 \documentclass[twocolumn]{pasj00}
\Accepted{accepted for publication in PASJ}
\SetRunningHead{Radiation Drag in Black Hole Outflows}{H. R. Takahashi
\& K. Ohsuga}

\usepackage{times}
\usepackage{bm}
\usepackage{natbib, aas_macros}
\def\bmath#1{\mbox{\boldmath $#1$}}
\begin{document}

\title{Radiation Drag Effects in Black Hole Outflows from Super-critical
Accretion Disks via Special Relativistic Radiation Magnetohydrodynamics Simulations}

\author{Hiroyuki R. Takahashi%
}
\affil{%
Center for Computational Astrophysics, National
  Astronomical Observatory of Japan, Mitaka, Tokyo 181-8588, Japan
}
\email{takahashi@cfca.jp}
\author{Ken Ohsuga%
}
\affil{%
Division of Theoretical Astronomy, National
  Astronomical Observatory of Japan, Mitaka, Tokyo 181-8588, Japan\\
School of Physical Sciences,Graduate University of
Advanced Study (SOKENDAI), Shonan Village, Hayama, Kanagawa 240-0193, Japan
}

\maketitle

\begin{abstract}
 By performing 2.5-dimensional special relativistic radiation 
 magnetohydrodynamics simulations,
 we study the super-critical accretion disks and 
 the outflows launched via the radiation force.
 We find that the outflow is accelerated by the radiation flux force,
 but the radiation drag force prevents the outflow velocity
 from increasing.
 The outflow velocity saturates around $30-40 \%$ 
 of the light speed around the rotation axis,
 since then the flux force balances with the drag force.
 Our simulations show that the outflow velocity 
 is kept nearly constant
 in the regime of $\dot{M}_{\rm BH}\sim 10^{2-3}L_{\rm Edd}/c^2$,
 where $\dot{M}_{\rm BH}$ is the mass accretion rate,
 $L_{\rm Edd}$ is the Eddington luminosity,
 and $c$ is the light speed.
 Such a faster outflow is surrounded by a 
 slower outflow of $\sim 0.1 c$.
 This velocity is also determined by 
 force balance between the radiation flux force and the radiation drag.
 The radiation drag works to collimate 
 the slower outflow in cooperation with the Lorentz force,
 although the faster outflow is mainly collimated by the
 Lorentz force.
 The kinetic energy is carried by the slower outflow
 rather than by the faster outflow.
 The total kinetic luminosity of the outflow
 as well as the photon luminosity is $\sim L_{\rm Edd}$,
 almost independent of the mass accretion rate.

\end{abstract}

\section{Introduction}
A black hole accretion disk system is one of the most energetic
phenomena in the universe. A mass accretion onto the black holes results
in effective release of gravitational energy. 
According to the mass accretion rate,
three accretion modes, i.e.,
standard disk model, 
radiatively inefficient accretion flow model, 
and slim disk model have been proposed by 
\cite{1973A&A....24..337S}, \cite{1977ApJ...214..840I},
\cite{1988ApJ...332..646A}, and \cite{1994ApJ...428L..13N}
\citep[see also][for a review]{2008bhad.book.....K}.
Whereas above three models are established by
one-dimensional approach, 
multi-dimensional hydrodynamic and magnetohydrodynamic (MHD) numerical
simulations of the accretion disks 
have been performed.
In MHD simulations, the phenomenological viscosity model
is not used, since 
the disk viscosity is magnetic origin, i.e.,
magnetorotational instability \citep{1991ApJ...376..223H}. 
However, the radiation transfer should be solved 
in order to investigate the luminous accretion modes.

By 2.5-dimensional radiation magnetohydrodynamics (RMHD) simulations,
\cite{2009PASJ...61L...7O} for the fist time 
succeeded in reproducing three accretion modes by one numerical code
\citep[see also][]{2011ApJ...736....2O}.
They revealed that the super-critical accretion,
of which the mass accretion rate
is over the critical rate ($L_{\rm E}/c^2$),
shines at the super-Eddington luminosity,
where $L_{\rm E}$ is the Eddington luminosity
and $c$ is the light speed.
In this case, 
since the huge amount of photons is mainly released towards 
the rotation axis of the disks, 
the radiation force does not prevent the mass accretion along the disk
plane \citep{2007ApJ...670.1283O}.
From the surface of the super-critical accretion disks,
powerful jets or outflows are launched by the strong radiation force.
\cite{2010PASJ...62L..43T} showed that 
the jets are accelerated by the radiation force and 
collimated by the Lorentz force.
This type of jet seems to explain mildly relativistic,
powerful jet from the microquasar, SS 433. 
However, their simulations are not fully taken 
account of the relativistic effect,
though the maximum velocity of the jet is 
several $10 \%$ of the light speed.

The highly relativistic jets,
of which the velocity is close to the light speed,
are thought to associate with the black hole accretion flows;
e.g., microquasar GRS 1915+105 \citep{1994Natur.371...46M,1999MNRAS.304..865F}, 
active galactic nuclei \citep{1995ApJ...447..582B,2012A&A...538L..10G},
and gamma-ray bursts \citep{2009Sci...323.1688A,2009ApJ...702..489R}.
The relativistic effects should play important roles
for such highly relativistic flows.
For instance, the radiation drag force
decelerates the outflows
in contrast with the acceleration via the radiation flux force.
Thus, for a non-relativistic approach,
the outflow velocity should be overestimated.
The relativistic RMHD simulations are required
to study the radiatively-driven high-velocity outflows.
Recently, special relativistic (SR)
\citep{2013ApJ...764..122T,2013ApJ...772..127T} 
and general relativistic (GR)
\citep{2008PhRvD..78b4023F, 2011MNRAS.417.2899Z, 
  2012MNRAS.426.1613R, 2013MNRAS.429.3533S}
RMHD code has been developed,
and GR-RMHD simulation 
of the super-critical disks are initiated 
\citep{2014MNRAS.441.3177M, 2014MNRAS.439..503S,2014arXiv1407.4421S}.

In this paper, 
we perform 2.5-dimensional SR-RMHD 
simulation of the super-critical accretion disks and launching outflows.
For the outflows, 
we investigate the deceleration via the radiation drag 
as well as the the acceleration via the radiation flux force.
The terminal velocity is determined by the balance
between above two forces.
This paper is organized as follows. In Section\ \ref{sec:equation}, we
introduce basic equations, and describe initial and boundary conditions. 
We show global inflow-outflow structure and detailed analysis of 
acceleration/deceleration of the outflow in Section\ \ref{sec:results}. 
Lastly, Section\ \ref{sec:summary} is devoted to
conclusions and discussion.

\section{Basic Equations, Initial and Boundary Conditions}\label{sec:equation}
We solve a full set of SR-RMHD equations.
Here, the Greek suffixes, $\mu$ and $\nu$, take values of
0, 1, 2, and 3, while the Latin suffixes of $i$, $j$, and $k$
take values of 1, 2, and 3.
By taking light speed $c$ as unity hereafter,
the basic equations of ideal
magnetofluids consist of mass conservation equation
\begin{equation}
 \partial_\nu (\rho u^\nu)=0,
  \label{eq:mass}
\end{equation}
energy-momentum conservation,
\begin{eqnarray}
& \partial_\nu 
  \left[
   \left(w_{\rm g} + \frac{b^2}{4\pi}\right)u^\mu u^\nu 
   - \frac{b^\mu b^\nu}{4\pi}
   + \left(p_{\rm g} + \frac{b^2}{8\pi}\right)\eta^{\mu\nu}
  \right]\nonumber \\
&=G_{\rm rad}^\mu + f_{\rm grav}^\mu,
  \label{eq:mhdmom}
\end{eqnarray}
and induction equation,
\begin{equation}
 \partial_\mu (u^\mu b^\nu - u^\nu b^\mu) = 0,
\label{eq:induction}
\end{equation}
where 
$\rho$ is the proper mass density, 
$p_{\rm g}$ is the gas pressure,
$u^\mu[=\gamma(1, v^j)]$ is four velocity,
$\gamma (=1/\sqrt{1-v_i v^i}$) is the Lorentz factor,
and $\eta^{\mu\nu}$ is the Minkowski metric
(of which the signature is [$-, +, +, +$]
in the present paper).
By supposing a simple $\Gamma$-law polytropic equation of state,
the gas enthalpy, $w_{\rm g}$, is given by
\begin{equation}
 w_{\rm g} = \rho + \frac{\Gamma}{\Gamma - 1} p_{\rm g},
\label{enthalpy}
\end{equation}
with $\Gamma$ being assumed to be $5/3$ throughout the present study.
A covariant magnetic field $b^\mu$ is related to the
magnetic field in laboratory frame, $B^j$, as
\begin{equation}
 b^\mu = \left[u_i B^i,
	  \frac{B^j + (u_i B^i)u^j}{\gamma}
	 \right],
 \label{eq:b}
\end{equation}
and an external force is described as
$f_{\rm grav}^\mu =-\gamma^2w_{\rm g}(u^i \partial_i \psi, \partial_i \psi)$,
where $\psi=-GM_{\rm BH}/(r-r_{\rm S})$ is the Pseudo-Newtonian
potential \citep{1980A&A....88...23P}.
Here, $M_{\rm BH}$ is the black hole mass,
$r$ is the distance from the central black hole,
and $r_{\rm S}(=2GM_\mathrm{BH})$ is the Schwarzschild radius.
In the present paper, we set $M_{\rm BH}$ to be $10 M_\odot$.

The radiation four force, $G_{\rm rad}^\mu$, is given by
\begin{eqnarray}
&& G^0_{\rm rad}= -\rho \kappa_\mathrm{a}
  \left(4\pi \gamma \mathrm{B} - \gamma E_{\rm rad} 
+ u_i F_{\rm rad}^i \right)
 \nonumber \\
&& - \rho \kappa_\mathrm{s}
\left[\gamma (\gamma^2-1) E_{\rm rad} + \gamma u_j u_k P^{jk}_\mathrm{rad}
-\left(2\gamma^2 - 1\right) u_i F_{\rm rad}^i \right],\label{geq:G0}
\end{eqnarray}
and 
\begin{eqnarray}
 G_{\rm rad}^{j} &=&- 4\pi \rho \kappa_\mathrm{a} \mathrm{B} u^j
 + \rho (\kappa_\mathrm{a} + \kappa_\mathrm{s})(\gamma F_{\rm rad}^j-u_kP^{jk}_\mathrm{rad}) 
 \nonumber \\
 &&-\rho \kappa_\mathrm{s} u^j
\left(\gamma^2 E_{\rm rad} - 2\gamma u_k F^k_\mathrm{rad} + u_k u_l P_{\rm rad}^{kl}\right),\label{geq:Gi}
\end{eqnarray}
where
$\kappa_\mathrm{a}=6.4\times 10^{22} \rho T_g^{-3.5} \ \mathrm{cm^2\ g^{-1}}$ and
$\kappa_\mathrm{s}=0.4 \ \mathrm{cm^2\ g^{-1}}$ 
are the Rosseland mean free-free absorption coefficient and the electron
scattering coefficient measured in the comoving frame,
$E_{\rm rad}$ is the radiation energy density,
$F_{\rm rad}^i$ is the radiation flux,
$P_{\rm rad}^{ij}$ is the radiation stress tensor,
$\mathrm{B}$ is the blackbody intensity,
and $T_g$ is the gas temperature.

The radiation energy density and the radiation flux are 
solved with using the zero-th and first order moment equations of
\begin{equation}
 \partial_t E_{\rm rad} + \partial_j F_{\rm rad}^j = - G_{\rm rad}^0,
  \label{eq:rade}
\end{equation}
and
\begin{equation}
 \partial_t F_{\rm rad}^j + \partial_i P_{\rm rad}^{ij} = - G_{\rm rad}^j.
  \label{eq:radf}
\end{equation}
The blackbody intensity $\mathrm{B}$ is described by gas
temperature $T_{\rm g}$ by
 \begin{equation}
 \mathrm{B}= \frac{a_{\rm rad} T_{\rm g}^4}{4\pi},\label{geq:bb}
 \end{equation}
here $a_{\rm rad}$ is the radiation constant,
and the gas temperature is
determined by the Boyle--Charles's law:
\begin{equation}
 p_{\rm g} =  \frac{\rho k_{\rm B} T_{\rm g}}{\mu m_{\rm p}},\label{geq:eos}
\end{equation}
where $k_{\rm B}$ is the Boltzmann constant,
$m_{\rm p}$ is the proton mass,
and $\mu(=0.5)$ is a mean molecular weight.
As a closure, we adopt M-1 closure in which 
the Eddington tensor $D^{jk}(\equiv P_{\rm rad}^{jk}/E_{\rm rad})$, 
is expressed as 
\begin{equation}
 D^{jk}=\frac{1-\chi}{2}\delta^{jk} + \frac{3\chi-1}{2}n^j n^k,
  \label{geq:DtensM1}
\end{equation}
where 
\begin{equation}
 \chi = \frac{3 + 4 |\bmath f|^2}{5 + 2 \sqrt{4 - 3 |\bmath f|^2}},
\end{equation}
\begin{equation}
 f^j = \frac{F_{\rm rad}^j}{E_{\rm rad}},
\end{equation}  
and
\begin{equation}
 n^j = \frac{F_{\rm rad}^j}{|\bmath F_{\rm rad}|}. \label{geq:DtensM1end}
  \label{M1D}
\end{equation}

We assume the axisymmetric $(\partial_\phi = 0)$ and reflecting boundary
at $\theta =0, \pi$.
At $\theta =0$ and $\pi$, $\rho$, $p_{\rm g}$, $v^r$, $B^r$, $E_{\rm rad}$, and
$F_{\rm rad}^r$ are symmetric, while others are antisymmetric.
For the inner ($r = 2 r_{\rm S}$) and outer ($r=534 r_{\rm S}$) boundaries, 
we use free boundary conditions and allow for matter 
and the radiation to go out but not to come in.
If the radial components of the velocity 
and the radiative flux are positive (negative)
at the innermost (outermost) grid, they are set to be zero.
For the magnetic fields,
we also employ the free boundary condition for the tangential components, 
$B^\theta$ and $B^\phi$, 
while the normal component, $B^r$, is determined to
satisfy $\nabla \cdot \bmath B = 0$. 

We start simulation with a low density, non-rotating, 
and non-magnetized corona surrounding
the black hole. The coronal density is given by 
\begin{eqnarray}
 \rho = \rho_{\rm c}
  \exp
  \left\{
   -\frac{\mu m_{\rm p}}{k_{\rm B} T_{\rm c}}
   \left[\psi(r) - \psi(r=100 r_{\rm S})\right]
  \right\}.
\end{eqnarray}
Here $T_{\rm c} = 10^{12}\mathrm{K}$ is the coronal temperature 
and $\rho_{\rm c} = 10^{-8}\ \mathrm{g\ cm^{-3}}$. 
The radiation temperature,
$T_{\rm rad}\equiv (E_{\rm rad}/a_R)^{1/4}$,
is uniform as $T_{\rm rad} =10^5\ \mathrm{K}$.

Following \cite{2003ApJ...592.1042I}, 
we continuously inject gas inside a torus,
which is located on equatorial plane 
and surrounding the black hole.
The curvature radius of the torus and 
the radius of the torus tube 
are $R_{\rm torus}=80 r_{\rm S}$ and $r_{\rm torus}=15 r_{\rm S}$, respectively.
In the torus, an increment of the gas density per unit time is
$\dot \rho_\mathrm{inj} 
= \dot M_{\rm inj}/(2\pi^2 r_{\rm torus}^2 R_{\rm torus})$, 
where $\dot M_{\rm inj}$ is the mass injection rate
and set to be $10^5L_{\rm E}$.
The temperature and the angular momentum of the injected matter
are set to be $10^{10}$K and 
be equal to the Keplerian angular momentum at 
$r=R_{\rm Kep}=60r_{\rm s}$.
%
%
In the torus, we also inject the poloidal magnetic field at each time step,
which is given by the increment of the azimuthal component 
of the vector potential as
\begin{eqnarray}
 \Delta A^\phi &=& \sqrt{\frac{8\pi \Delta \rho c_{s,\mathrm{inj}}^2}{\beta_{\rm inj}}}
  \frac{r_{\rm torus} R_{\rm torus}}{r \sin\theta}\nonumber \\
 &\times &\exp 
  \left[-\frac{8(r^2 + R_{\rm torus}^2 + 2 r R_{\rm torus} \sin\theta)}
   {r_{\rm torus}^2}\right],
\end{eqnarray}
where $\Delta \rho=\dot \rho_\mathrm{inj}\Delta t$ and $c_{s,{\rm inj}}$ are
the increase of the mass density inside the torus within the time step,
$\Delta t$, and the sound velocity of the
injected gas, respectively. We set $\beta_\mathrm{inj}=100$. 
We compute the plasma-$\beta$ inside the torus at each time step and
suspend injection of the magnetic field 
when plasma-$\beta$ is smaller than 30 for numerical stability
\citep{2008ApJ...677..317I}.

We have to note that our simulation has been performed in 2-dimension by 
assuming axisymmetry, so that the anti-dynamo effect works and the magnetic 
field cannot be maintained \citep{1933MNRAS..94...39C}. So as to avoid
this problem, we inject the gas which has the poloidal magnetic field,
realizing a steady accretion in the present work (see, section
\ref{sec:overview}). 
Our procedure would not be unnatural, since the magnetized matter is
thought to be transported from outer region to the disks in
reality. Also, the phenomenological dynamo 
model is employed to amplify the magnetic fields in some of 2-dimensional 
simulations 
\citep{2013MNRAS.428...71B, 2014MNRAS.440L..41B, 2014arXiv1407.4421S}.
The 3-dimensional simulations, in which the anti-dynamo 
does not work, are performed by \cite{2014MNRAS.441.3177M}. The impact of our 
boundary condition on the resulting accretion flow structure should be 
verified by comparing results between these different models.

\section{Results}\label{sec:results}
\subsection{Overview}\label{sec:overview}
After simulation starts, a gas is injected in the torus. 
Since the injected gas is not in dynamically equilibrium 
and since a gas pressure is larger in the torus than in the corona, 
the injected gas expands by the gas pressure gradient force at once.
This gas falls back towards the equatorial plane 
and accumulates around $r=R_\mathrm{Kep}$. 
Such an expansion behavior is remarkable only 
at the elapse time of $t\lesssim 1$ s.
%
After a few second, a MHD turbulence develops 
and gas starts to accrete onto the black hole. 
The mass accretion rate onto the black hole 
rapidly increases at $t=2.4\ \mathrm{s}$.
After that, 
the gas is continuously swallowed by the black hole
at the super-critical rate, $\gg L_{\rm E}$.
Then, the radiation energy is enhanced in the accretion disks
since a part of the gas internal energy is converted 
through the free-free emission.
The strong radiation force drives the outflows from the disks.
At that time, 
since the torus is embedded in the dense matter,
the expansion behavior does not occur.
Hence, the incipient expansion 
does not affect the accretion motion and launching outflows.

\begin{figure}
 \includegraphics[width=8cm]{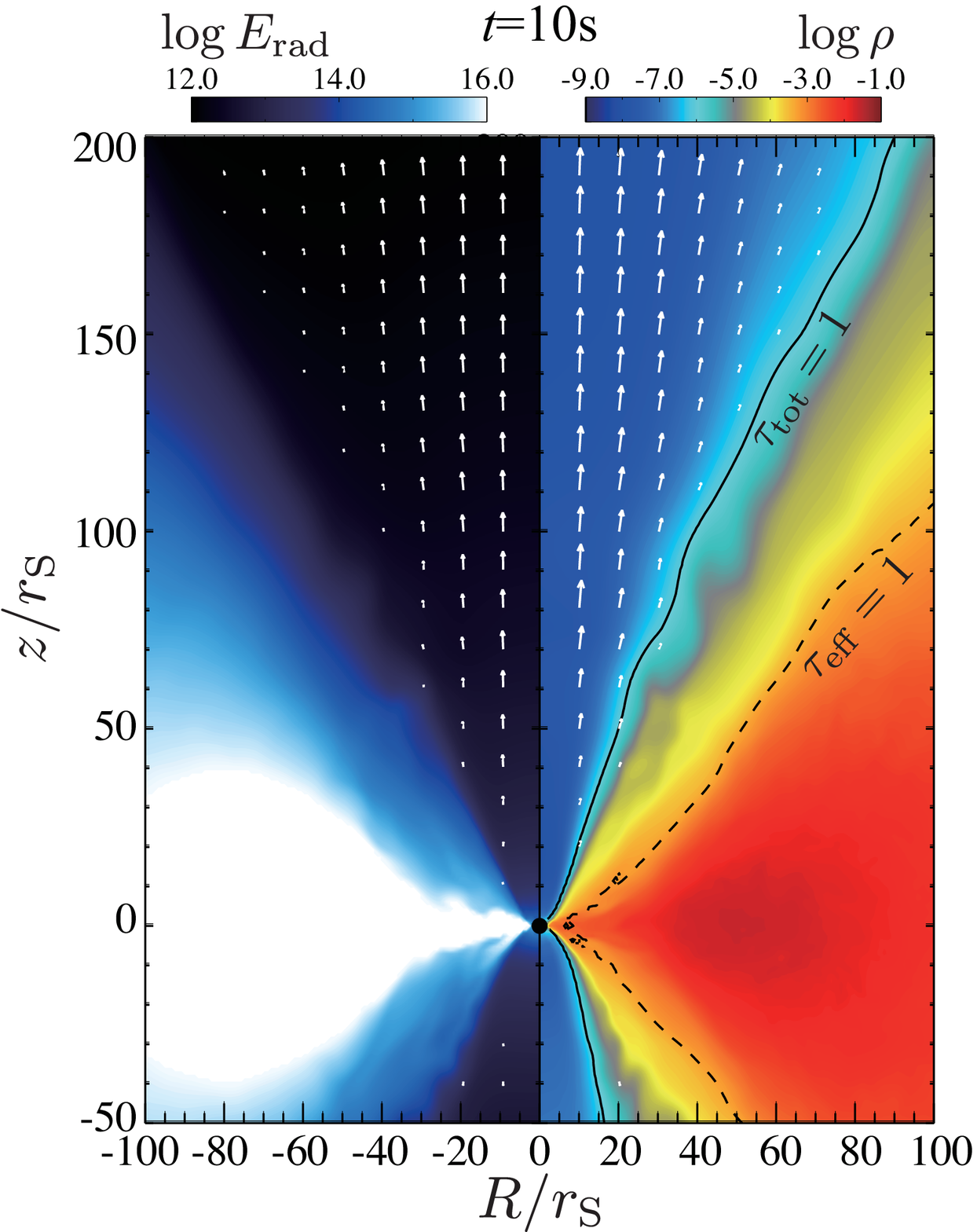}
  \caption{Global structure of radiation dominated accretion disks 
 and launching outflows at $t=10\ \mathrm{s}$.
 Color shows $E_{\rm rad}$ and $\rho$ and arrows indicate 
 $\bmath F_{\rm rad}/E_{\rm rad}$
 and $\bmath v$ for left and right panels, respectively.
 Solid and dashed curves in the right panel show $\tau_\mathrm{tot}=1$ and 
 $\tau_\mathrm{eff}=1$, respectively.
 }
\label{fig:fig1}
\end{figure}

In the left panel of Figure \ref{fig:fig1}, 
we show the distribution of the radiation energy density (color)
and the radiation flux normalized by the radiation energy density
is denoted by arrows in $R$-$z$ plane.
Here $R(=r \sin\theta)$ and $z(=r \cos\theta)$ 
are horizontal and vertical distances.
A right panel indicates the density distribution (color)
overlaid with the velocity vectors.
Solid and dashed curves in the right panel show $\tau_{tot}=1$ and 
$\tau_{eff}=1$, respectively. Here $\tau_{tot}$ and $\tau_{eff}$ are the total and 
effective optical depths integrated from $z=\pm 300 r_\mathrm{S}$ to $z=0$
in $\mp z$-direction.
We find in the right panel that
the geometrically thick disk forms (red).
This disk is supported by the strong radiation pressure.
Indeed, it is found that
the radiation energy density is enhanced in the disk.
Since the disk is very optically thick for the electron scattering,
most of the radiation energy is trapped inside the accretion disks.

In contrast of $|F_{\rm rad}|\ll E_{\rm rad}$ in the disk,
the radiation energy is effectively transported above the disk
($z/R\gtrsim 2$)
due to small density (small optical depth).
Especially, we find $|F_{\rm rad}|\sim E_{\rm rad}$ around the rotation axis,
meaning that the photons freely move in vertical direction.
Then, the radiation flux force effectively 
accelerates the gas, leading to the outflows.
Such an outflowing motion is clearly shown
in the right panel (see velocity vectors above the disk).

\begin{figure}
 \includegraphics[width=8cm]{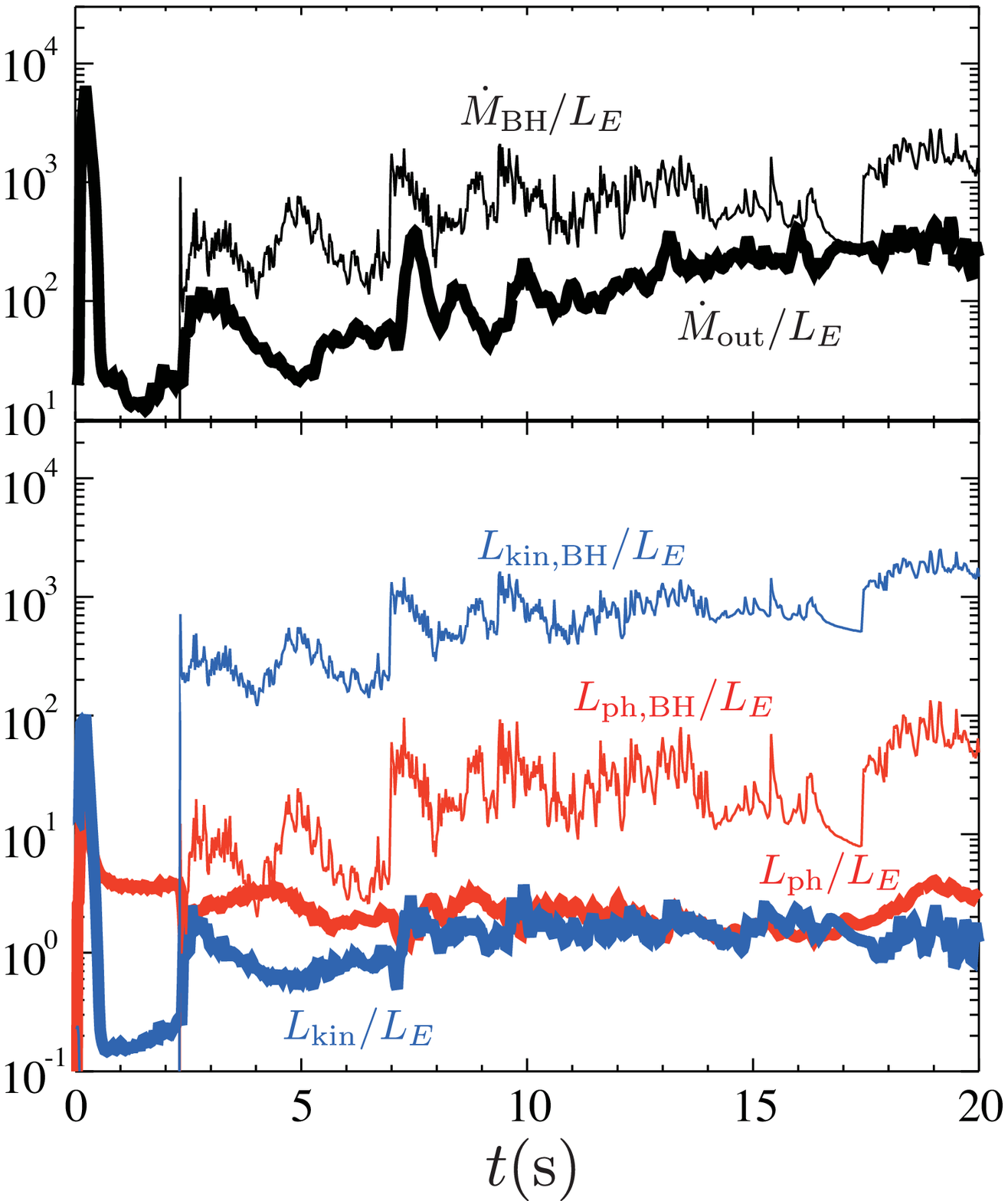}
  \caption{Time evolution of 
 the mass accretion rate onto the black hole 
 ($\dot{M}_{\rm BH},$ thick black),
 the mass outflow rate 
 ($\dot{M}_{\rm out}$, thin black),
 the photon luminosity ($L_{\rm ph}$, thick red),
 and the kinetic power ($L_{\rm kin}$, thick blue).
 Red and blue thin curves denote 
 the photon luminosity and kinetic power 
 swallowed by the black hole 
 ($L_{\rm ph, BH}$ and $L_{\rm kin, BH}$), respectively.}
\label{fig:fig2}
\end{figure}

We plot in Figure \ref{fig:fig2} 
a time evolution of the mass accretion rate
onto the black hole (thin black),
\begin{equation}
 \dot M_\mathrm{BH} \equiv 
  -2\pi (2r_{\rm S})^2
  \int_{-\pi}^{\pi} 
  \gamma \rho v^r \sin\theta
  d\theta,
  \label{eq:mdotin}
\end{equation}
and the mass outflow rate (thick black),
\begin{eqnarray}
 \dot M_\mathrm{out} 
  &\equiv& \left. 2\pi \int_0^{150r_{\rm S}}
  \gamma \rho v^z  H(v^r - v_\mathrm{esc}) 
  R dR\right|_{z=300r_\mathrm{S}}\nonumber \\
 &-&  \left. 2\pi \int_0^{150r_{\rm S}}
  \gamma \rho v^z  H(v^r - v_\mathrm{esc}) R dR
\right|_{z=-300r_\mathrm{S}}.
  \label{eq:mdotout}
\end{eqnarray}
$H$ is the Heaviside function,
\begin{equation}
 H(x) =\left\{\begin{array}{ll}
	0  & ( x < 0 ) \\
	1  & ( x > 0 )
       \end{array}\right.
\end{equation}
and $v_\mathrm{esc}=\sqrt{r_\mathrm{S}/r}$ is the escape velocity.
Here we ignored relativistic effects since the outflow velocity is only
mildly relativistic (see, Section\ \ref{sec:outflow_property}) and the rest
mass is the dominant energy density.
The photon luminosity, $L_{\rm ph}$ (thick red),
the kinetic power, $L_{\rm kin}$ (thick blue),
the photon luminosity swallowed by the black hole, $L_{\rm ph, BH}$
(thin red), 
and the kinetic power swallowed by the black hole, $L_{\rm kin, BH}$
(thin blue)
are also represented in this figure.
These values are evaluated as 
\begin{eqnarray}
  L_{\rm ph} &\equiv &
  \left. 2\pi \int_0^{150r_{\rm S}} F_{\rm rad}^z R dR
   \right|_{z=300r_\mathrm{S}}\nonumber \\
&-& \left. 2\pi \int_0^{150r_{\rm S}} F_{\rm rad}^z R dR
   \right|_{z=-300r_\mathrm{S}},
   \label{eq:Lrout}
\end{eqnarray}
\begin{eqnarray}
  L_{\rm kin} &\equiv &
   \left. 2\pi \int_0^{150r_{\rm S}} 
  \gamma (\gamma w_{\rm g} - \rho) v^z H(v^r - v_\mathrm{esc}) R dR
  \right|_{z=300r_\mathrm{S}}
   \nonumber \\
&-&  \left. 2\pi \int_0^{150r_{\rm S}} 
  \gamma (\gamma w_{\rm g} - \rho) v^z H(v^r - v_\mathrm{esc}) R dR
  \right|_{z=-300r_\mathrm{S}},
  \label{eq:Lkout}
\end{eqnarray}
\begin{equation}
  L_{\rm ph, BH} \equiv  -2\pi (2r_{\rm S})^2
   \int_{-\pi}^{\pi} F_{\rm rad}^r \sin\theta d\theta,
   \label{eq:Lrin}
\end{equation}
and
\begin{equation}
 L_{\rm kin, BH} \equiv -2\pi (2r_{\rm S})^2
   \int_{-\pi}^{\pi} \gamma (\gamma w_{\rm g} - \rho) v^r 
   \sin\theta d\theta.
   \label{eq:Lkin}
\end{equation}

\begin{figure}
 \includegraphics[width=8cm]{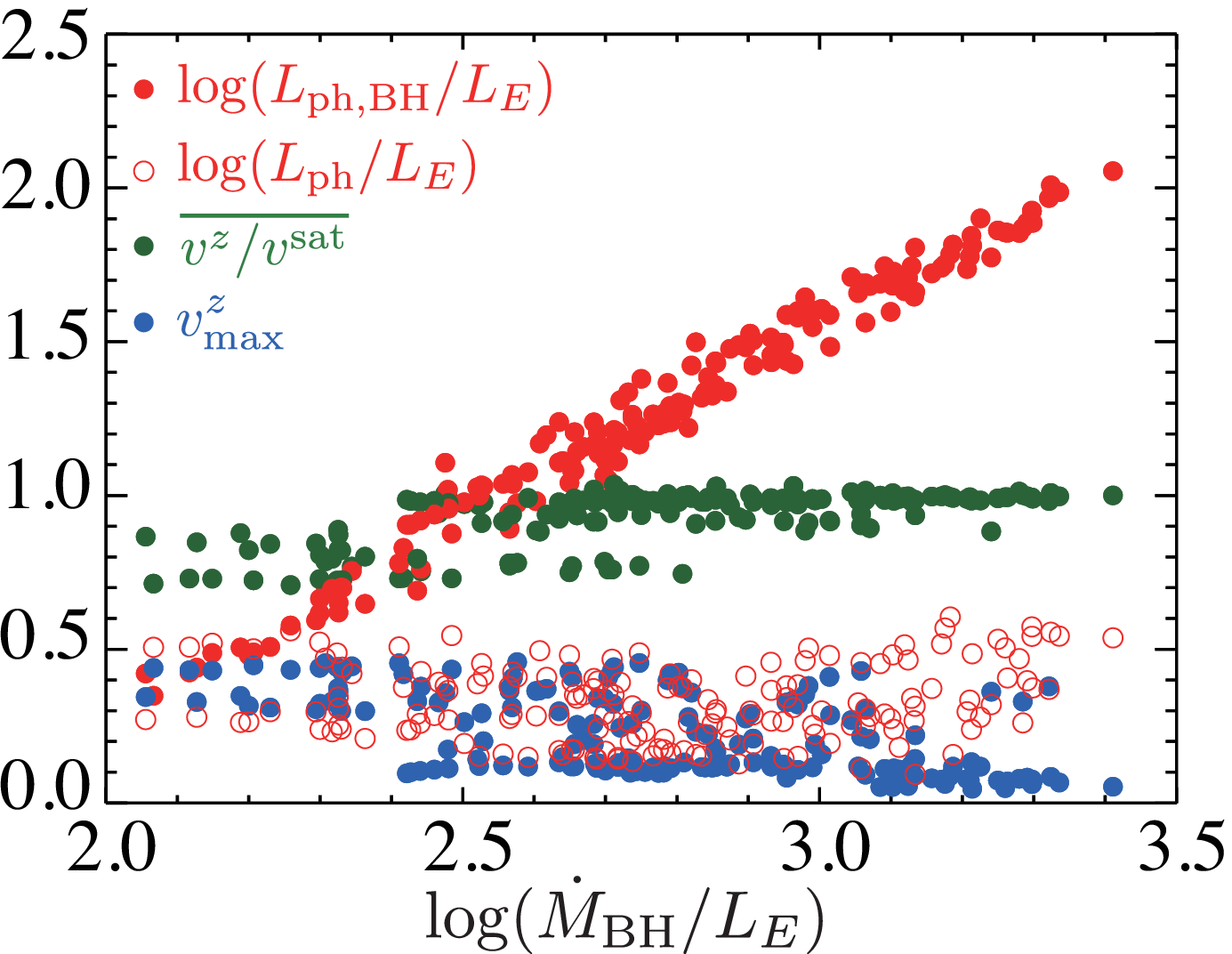}
  \caption{The photon luminosity ($L_{\rm ph}$, open red circle),
 the photon luminosity swallowed by the black hole
 ($L_{\rm ph, BH}$, filled red circle), 
 and the maximum outflow velocity ($v^z_{\rm max}$, blue circle)
 as a function of $\dot M_\mathrm{BH}$.
 Here, $v^z_{\rm max}$ is calculated at the
 surface of $z=\pm 300 r_{\rm S}$ and $R=[0,150 r_{\rm S}]$.
 We also plot  
 $\overline {v^z/v^\mathrm{sat}}$ 
 (green circle), where
 bar denotes the averaged value at the surface of 
 $z=\pm 300 r_{\rm S}$ and  $R=[0,150r_{\rm S}]$.}
\label{fig:fig3}
\end{figure}

After $t\simeq 3$ sec.,
we find that 
the mass accretion rate onto the black hole 
is about $\dot M_\mathrm{BH} \simeq 1000L_{\rm E}$
and slowly increases with time.
Due to the release of the gravitational energy via the mass accretion, 
the radiation energy is enhanced,
driving the outflow from the accretion disk. 
The outflow carries a large amount of gas 
at the rate of $\sim$ a few 10\% of the mass accretion rate, 
which exceeds the critical rate ($L_{\rm E}$). 
The kinetic power and the photon luminosity
are almost comparable to the critical rate, i.e.,
$L_{\rm kin} \sim L_{\rm ph} \sim L_{\rm E}$.
In addition, 
we find $L_{\rm kin}$ and $L_{\rm ph}$ are
much smaller than $L_{\rm kin, BH}$ and $L_{\rm ph, BH}$,
respectively.
This means that most of kinetic 
and radiation energies are swallowed by the black hole.
Since the huge amount of photons is swallowed by
the black hole with accreting gas (photon trapping),
the photon luminosity is not sensitive to the mass accretion rate
\citep{1978MNRAS.184...53B,2002ApJ...574..315O,2003ApJ...596..429O}.
This is clearly shown in Figure \ref{fig:fig3},
where $L_{\rm ph}$ (red open circle) as well as $L_{\rm ph, BH}$ (red 
filled circle) is 
plotted as a function of the accretion rate, ${\dot M}_{\rm BH}$.
In contrast with $L_{\rm ph}$,
we find $L_{\rm ph, BH}$ increases with an increase of 
${\dot M}_{\rm BH}$.
Such an effective photon trapping
is a characteristic feature in super-critical accretion disks 
\citep{2005ApJ...628..368O, 2007ApJ...670.1283O, 2011ApJ...736....2O}.

\begin{figure}
 \includegraphics[width=8cm]{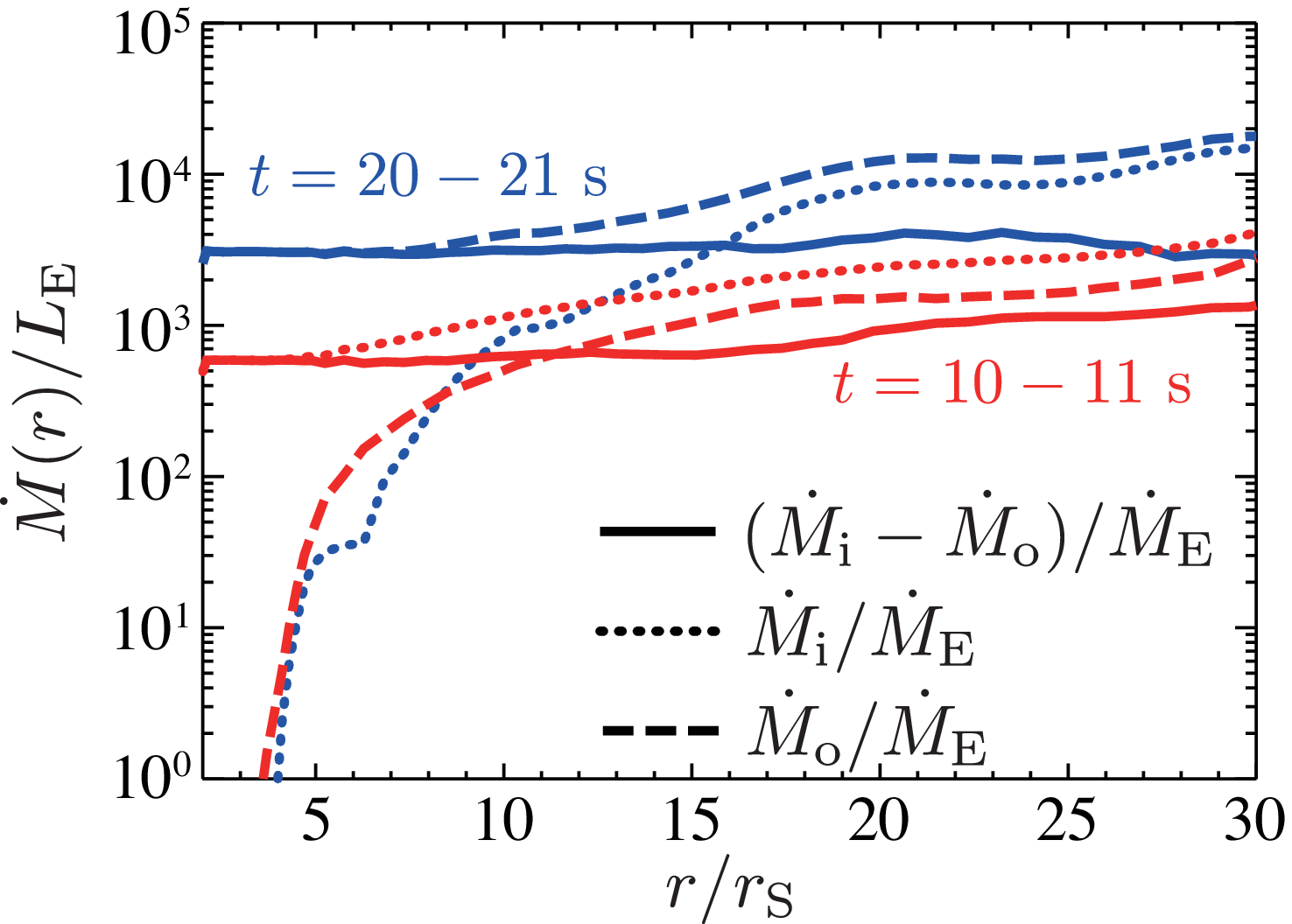}
  \caption{Radial profiles of accretion rate (dashed), outflow
 rate (dotted) passing through the spherical surface with polar radius
 $r$ averaged between
 $t=10-11\ \mathrm{s}$ (red) and $t=20-21\ \mathrm{s}$. The accretion
 rate is plotted by changing its sign. Solid curves denote for the total
 mass flow rate. }
\label{fig:fig4}
\end{figure}

We note that most of calculations in the previous study start from an 
initial torus, which is in hydrostatic equilibrium. In this method, the total 
mass within the computational domain decreases with time. In our simulations, 
the gas is injected at constant rate. The total mass in the simulation box
and the mass accretion rate onto the black hole increases with time. However, 
the disk is actually in steady state which is understood in Figure\
\ref{fig:fig4}. In this figure, we plot the mass inflow ($\dot
M_\mathrm{i}$ dashed) and outflow ($\dot M_\mathrm{o}$, dotted) 
rates as a function of r, 
\begin{equation}
 \dot M_\mathrm{i}(r) =-2\pi \int 
  \gamma \rho v^r H(-v^r) r^2 \sin\theta d\theta,
\end{equation}
and 
\begin{equation}
 \dot M_\mathrm{o}(r) = 2\pi \int 
  \gamma \rho v^r H(v^r)r^2 \sin\theta d\theta.
\end{equation}
Here note that $\dot M_\mathrm{i}$ is the same with $\dot M_\mathrm{BH}$
when we take $r=2 r_\mathrm{S}$. A total mass flow rate ($\dot
M_\mathrm{i}- \dot M_\mathrm{o}$) is also plotted by solid lines.
We can see that the inflow rate is larger than the outflow rate, and the total 
mass flow rate is almost constant in the region of
$r=[2-30]r_\mathrm{S}$. 
This indicates that the inflow equilibrium is attained, while the mass flow rate slowly 
increases with time. Thus, the results obtained by our simulation are not 
transient behavior but the quasi-steady structure for various mass accretion 
rate, $\dot M_\mathrm{BH}$.

Also, \cite{2014arXiv1407.4421S} showed that the mechanical energy
weakly increases with $\dot M_\mathrm{BH}$, while it is almost
independent of it in our results. Such a 
discrepancy might be due to the general relativistic effects and/or
$\dot M_\mathrm{BH}$. 
In \cite{2014arXiv1407.4421S}, the general relativistic effects are taken into 
consideration and the mass accretion rate is around $\lesssim 100 \dot
M_\mathrm{E}$. The present simulations are special relativistic version,
and the mass accretion exceeds $100 \dot M_\mathrm{E}$. Detailed study
of such a difference is left as an important future work.


%

\subsection{Outflow Properties}\label{sec:outflow_property}
Next, we focus on the outflow structure. 
We hereafter show time averaged values over $t=10-11\mathrm{s}$. 
Figure \ref{fig:fig5} shows 
vertical profiles of the outflow velocity, $v^z$ 
for $R=10 r_{\rm S}$(black solid), $30r_{\rm S}$(black dashed), 
and $60 r_{\rm S}$(black dotted).
We can see that a gas is accelerated at the small altitudes,
and its velocity finally saturates at the outer region
due to the radiation drag 
(we will discuss in next subsection).
In particular,
the outflow velocity for $R=10r_{\rm S}$ 
is proportional to the altitude at $\lesssim 80r_{\rm S}$,
and is kept constant in the region of $z \gtrsim 80r_{\rm S}$.
For $R=30r_{\rm S}$ and $60r_{\rm S}$,
the gas is accelerated up to $z\sim 240r_{\rm S}$ and $\sim 280r_{\rm S}$.
The terminal outflow velocity is larger near the rotation axis,
where the gas is blown away at the speed of $\sim 0.3$. 
The velocity for $R=60r_{\rm S}$ is slightly over 0.2
at a maximum.

Such $R$-dependence of the outflow velocity 
at $z=300 r_{\rm S}$ 
is clearly shown in Figure \ref{fig:fig6} (black solid).
As noted before, the faster outflow ($v^z\gtrsim 0.3$) is concentrated 
near the rotation axis ($R<30r_{\rm S}$), 
while slower outflow extends up to $R\sim 100 r_{\rm S}$. 
Its speed is typically $\sim 0.1-0.2$, 
which exceeds the escape velocity (dashed line). 
This structure is similar to spine-sheath structure
\citep{1989MNRAS.237..411S, 2003NewAR..47..667M}, 
in which the faster outflow/jet 
is surrounded by the slower outflow.

In this figure, 
we also plot $R$-dependent kinematic power ($\Delta L_{\rm kin}$, blue) 
and photon luminosity ($\Delta L_{\rm ph}$, red).
They are assessed as
\begin{eqnarray}
  \Delta L_{\rm kin} &\equiv& 2\pi 
   \left. 
    \int_{R-\Delta R/2}^{R+\Delta R/2}
   \gamma (\gamma w_{\rm g} - \rho) v^z H(v^r - v_\mathrm{esc}) R dR
   \right|_{z=300r_\mathrm{S}} \nonumber \\
 &-& 2\pi \left. \int_{R-\Delta R/2}^{R+\Delta R/2}
   \gamma (\gamma w_{\rm g} - \rho) v^z H(v^r - v_\mathrm{esc}) R dR
   \right|_{z=-300r_\mathrm{S}},
   \label{eq:dLkin}
\end{eqnarray}
and
\begin{eqnarray}
  \Delta L_{\rm ph} &\equiv &
   \left. 
    2\pi    \int_{R-\Delta R/2}^{R+\Delta R/2}
   F_{\rm rad}^z R dR
   \right|_{z=300 r_\mathrm{S}}\nonumber \\
 &-& 
   \left. 
    2\pi    \int_{R-\Delta R/2}^{R+\Delta R/2}
   F_{\rm rad}^z R dR
   \right|_{z=-300 r_\mathrm{S}},
   \label{eq:dLph}
\end{eqnarray}
where $\Delta R \equiv 2 r_\mathrm{S}$.
We find that the kinetic energy is mainly carried 
by the slower outflow. 
The relation of $\Delta L_{\rm kin}\propto R$
implies that the kinetic energy flux
(the kinetic energy transported per unit surface and per unit time)
is almost constant within $R\sim 40r_{\rm S}$.
This figure also shows that such a kinetic energy flux
is larger near the rotation axis ($R\lesssim 40r_{\rm S}$) 
than in the region of $R\gtrsim 40r_{\rm S}$.
The profile of $\Delta L_{\rm ph}$ is similar with
that of $\Delta L_{\rm kin}$.
Thus, the vertical radiation flux is mildly 
collimated within $R\sim 40r_{\rm S}$.
However, due to a larger opening angle,
the kinetic power of the slow outflow
is dominant over that of the fast outflow,
and the radiation energy is mainly released 
from the region of $R\sim 40-100r_{\rm S}$.

\subsection{Radiation Drag Force}
In this subsection, 
we show effects 
of the radiation drag in the outflow.
Since the flow is quasi-steady, here, 
we introduce the steady state equation of motion,
\begin{eqnarray}
 \gamma^2 w_g \left( \bmath v \cdot \nabla \right) v^j
  = f_{\rm gas}^j + f_{\rm mag}^j + f_{\rm grav}^j 
 + f_{\rm rad}^j,
\end{eqnarray}
where
\begin{eqnarray}
  f_{\rm gas}^j &=& -\partial_j p_{\rm g},\label{eq:fp}\\
  f_{\rm mag}^j &=& \frac{1}{4\pi}
  \left[
    (\bmath v \times \bmath B)^j 
    \partial_i (\bmath v \times \bmath B)^i
  \right]
  +\frac{1}{4\pi} \left[
    (\nabla \times \bmath B)\times \bmath B 
  \right]^j,\label{eq:fm}\\
  f_{\rm rad}^j &=& f_{\rm rad-flux}^j + f_{\rm rad-drag}^j 
  + f_{\rm rad-corr}^j,\label{eq:fr}
\end{eqnarray}
are the gas pressure gradient, Lorentz, and radiation forces.
The radiation force consists of following three components;
the radiation flux force,
\begin{equation}
  f^j_{\rm rad-flux} = \gamma \rho (\kappa_\mathrm{a} + \kappa_\mathrm{s})F^j_{\rm rad},
  \label{eq:frf}
\end{equation}
the radiation drag,
\begin{equation}
f^j_{\rm rad-drag} =- \gamma \rho (\kappa_\mathrm{a} + \kappa_\mathrm{s})
 \left(E_{\rm rad} v^j + v_k P_{\rm rad}^{jk}\right),
  \label{eq:frd}
\end{equation}
and the relativistic correction,
\begin{equation}
  f^j_{\rm rad-corr} = \gamma \rho (\kappa_\mathrm{a} + \kappa_\mathrm{s})
  (v_i F^i_{\rm rad})v^j .
  \label{eq:frr}
\end{equation}
The radiation drag,
which is of the order of $v$,
works to slow down the relativistic flow.
The relativistic correction is ${\mathcal O}(v^2)$
so that it plays an important role only for 
highly relativistic flow.

\begin{figure}
 \includegraphics[width=8cm]{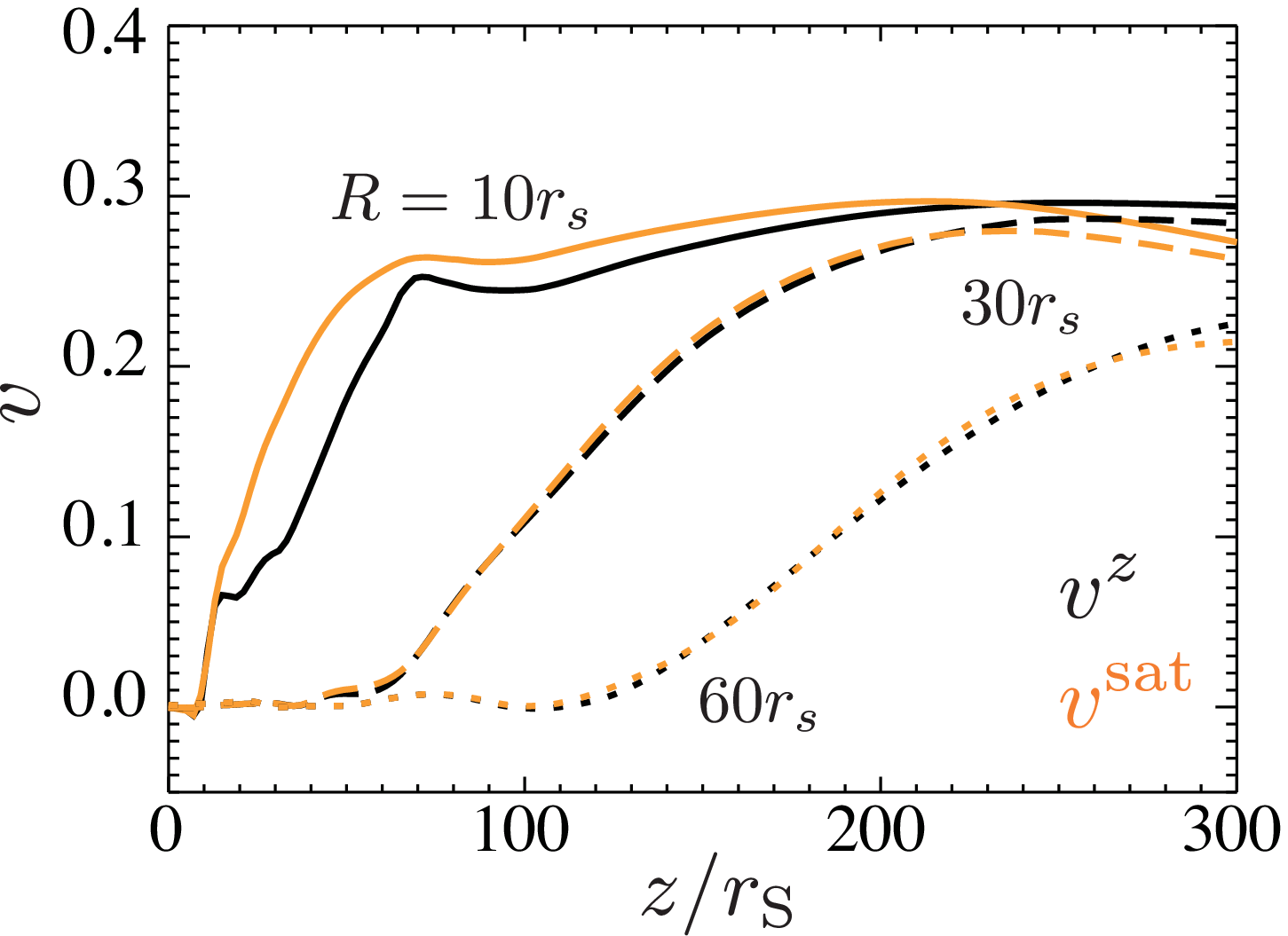}
 \caption{Vertical profiles of 
 the vertical component of the velocity ($v^z$, black) 
 and the saturation velocity ($v^{\rm sat}$, orange)
 for $R=10$ (solid), $30$ (dashed), $60 r_{\rm S}$ (dotted), respectively.}
\label{fig:fig5}
\end{figure}

\begin{figure}
 \includegraphics[width=8cm]{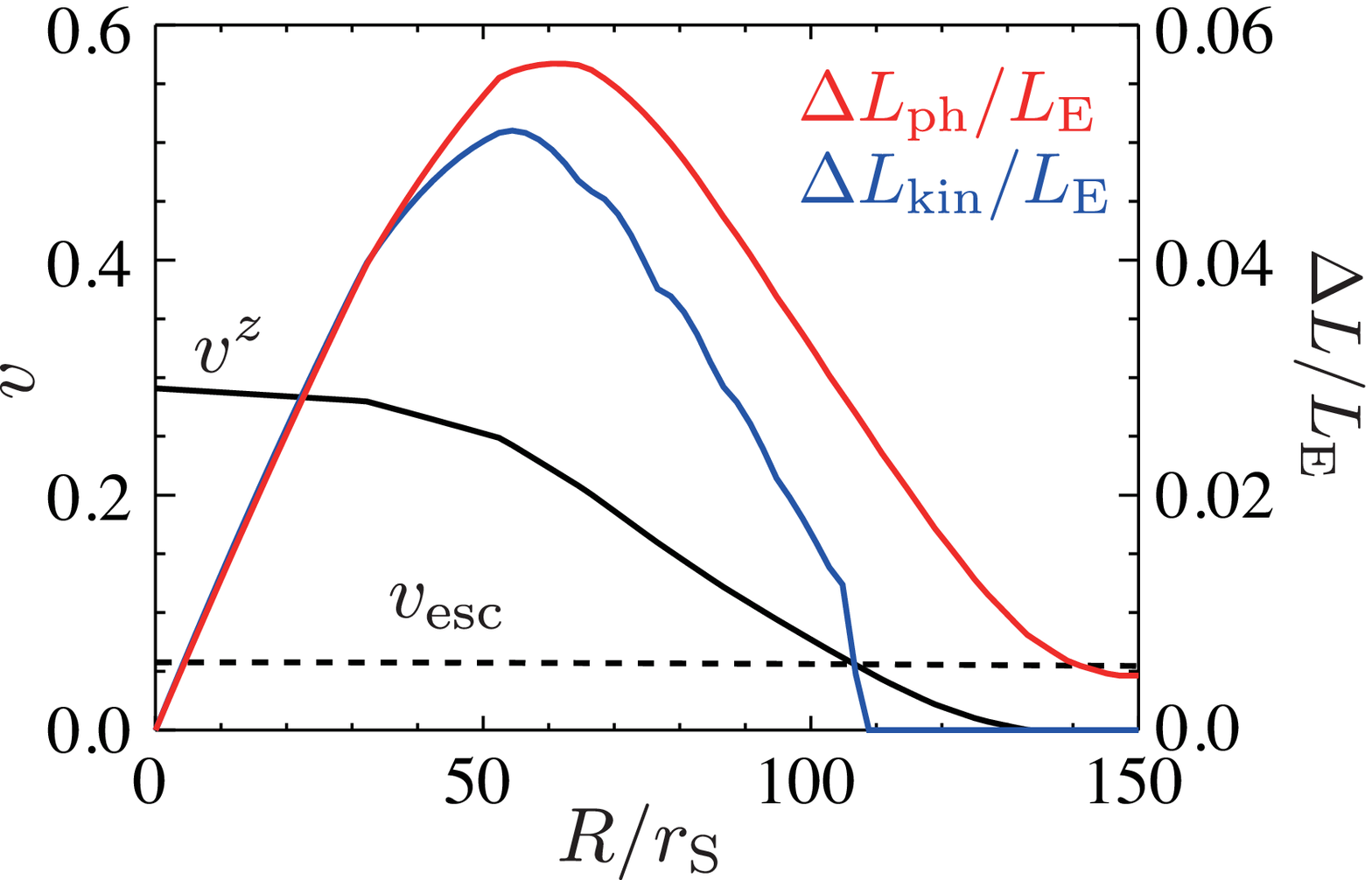}
 \caption{Radial profiles of the 
 vertical component of the velocity ($v^z$, black),
 $\Delta L_{\rm kin}$ (blue), and $\Delta L_{\rm ph}$ (red)
 at $|z|=300 r_{\rm S}$. 
 A dashed curve denotes for the escape velocity.}
 \label{fig:fig6}
\end{figure}

\begin{figure}
 \includegraphics[width=8cm]{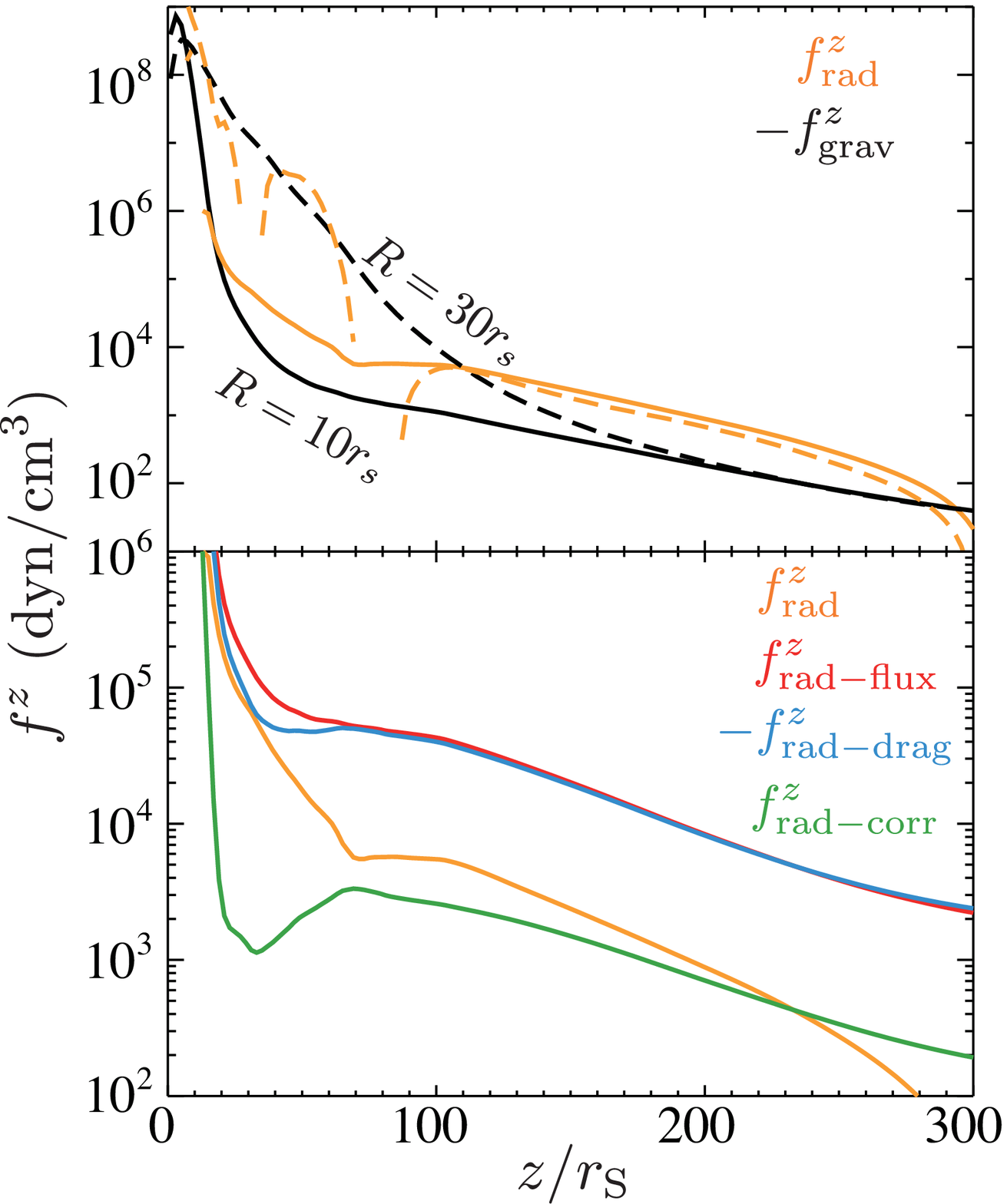}
  \caption{
 Vertical component of the force densities
 as a function of the altitude, $z$.
 Top panel: 
 the gravity ($-f^z_\mathrm{grav}$, black) 
 and the radiation force ($f^z_\mathrm{rad}$, orange) for 
 $R=10r_\mathrm{S}$ (solid) and $30 r_\mathrm{S}$ (dashed).
 Bottom panel: 
 the radiation flux force ($f^z_\mathrm{rad-flux}$, red),
 the radiation drag force ($-f^z_\mathrm{rad-drag}$, blue), 
 and the relativistic correction ($f^z_\mathrm{rad-corr}$, green) 
 at $R=10 r_\mathrm{S}$.
 The orange curve is the same as that of the top panel
 ($f^z_\mathrm{rad}$).}
\label{fig:fig7}
\end{figure}

In a top panel of Figure \ref{fig:fig7}, 
a vertical component of the radiation force (orange) 
and gravitational force (black) are plotted
along the vertical lines of 
$R=10 r_\mathrm{S}$ (solid) and $30 r_\mathrm{S}$ (dashed).
We can see that 
the radiation force tends to be weaker than the gravity
at small altitude (within the disk ), 
but exceeds the gravity above the disk.
The turn-around altitude is around
$20r_{\rm S}$ for $R=10r_{\rm S}$ and 
$110r_{\rm S}$ for $R=30r_{\rm S}$.
Although the gas pressure-gradient force and 
the Lorentz force are not plotted in this panel,
they are much smaller than the gravity.
Therefore, we conclude that 
the outflows are accelerated by the radiation force. 
The gas is mainly accelerated just above the turn-around altitude
since $f^z_{\rm rad}+f^z_{\rm grav}$ peaks there,
and the radiative acceleration becomes 
ineffective at large altitude. 
Note that 
the deviation between the radiation force
and the gravity is significant near the rotation axis.
This is the reason why 
the faster outflow/jet forms around the axis
(see Figures \ref{fig:fig1} and \ref{fig:fig5}).

Why the radiative acceleration decreases at large altitude?
This is due to the radiation drag.
A bottom panel of Figure \ref{fig:fig7} shows 
the vertical profile of $f^z_\mathrm{rad}$ (orange),
$f^z_{\rm rad-flux}$ (red), $-f^z_{\rm rad-drag}$ (blue) 
and $f^z_{\rm rad-corr}$ (green) at $R=10 r_{\rm S}$.
It is found that 
the $f^z_{\rm rad-flux}$ is much larger than the gravity at large altitude.
This implies 
that the radiation flux force continues to accelerate the gas.
However, 
the radiation drag, which is the downward force,
is comparable to the radiation flux force at $z\gtrsim 80r_{\rm S}$.
Thus, the radiative acceleration becomes inefficient at the large altitude.
In contrast, the radiation flux force is much larger than
the radiation drag force, $f^z_{\rm rad-flux} \gg -f^z_{\rm rad-drag}$,
just above the turn-around altitude
($z\sim 20 r_{\rm S}$),
leading to the effective acceleration.
Here we note that the outflow velocity is mildly relativistic 
so that $f^z_{\rm rad-corr}$ is negligible. 

The force balance of $f^z_{\rm rad-flux} \sim -f^z_{\rm rad-drag}$
is nearly equivalent to $F'^z_{\rm rad} \sim 0$, 
where $F'^z_{\rm rad}$ is the radiation flux measured in the comoving frame.
This is because that $F'^z_{\rm rad}$ is related to 
the radiation fields in the laboratory frame
as $F'^z_{\rm rad} \sim F^z - E_{\rm rad} v^z - v_i P_{\rm rad}^{zi}$,
and it is rewritten as 
$(f^z_{\rm rad-flux}-f^z_{\rm rad-drag})/[\rho (\kappa_\mathrm{a} + \kappa_\mathrm{s})]$.
From this fact, we can obtain the saturation velocity 
from $F'^z_{\rm rad}=0$ as
\begin{equation}
 v^{\rm sat}=\frac{F_{\rm rad}^z}{E_{\rm rad} + P_{\rm rad}^{zz}}.\label{eq:vsat}
\end{equation}
Here we assume $|v_z P^{zz}_\mathrm{rad}| \gg |v_r P^{rz}_\mathrm{rad}|, |v_\phi P^{\phi z}_\mathrm{rad}|$. 
In Figure \ref{fig:fig5}, we plot the saturation velocity by orange lines.
This figure clearly shows $v^z<v^{\rm sat}$ 
at $z\lesssim 220$ for $R=10 r_\mathrm{S}$.
In this region, 
the vertical component of the radiation flux in the comoving frame 
is positive and the gas is pushed in the vertical direction.  
We find $v^z\sim v^{\rm sat}$ above that region, 
implying that 
the radiation flux is almost zero 
in the comoving frame.
The radiation force thus cannot accelerate a gas further. 
In this figure, we can also see 
that $v^z$ is slightly less than $v^{\rm sat}$ 
at $z\lesssim 200 r_\mathrm{S}$ for $30r_{\rm S}$,
leading to the weak acceleration of the gas.

This radiation drag plays an important role 
when $\dot{M}_{\rm BH}\gg L_{\rm E}$.
In figure \ref{fig:fig3}, we show
a maximum outflow velocity, $v^z_{\rm max}$ (blue), 
which is computed at a surface of $|z|=300 r_{\rm S}$ and $R=[0,150r_{\rm S}]$. 
We also plot a ratio of saturation velocity 
and outflow velocity
averaged over the same surface for maximum outflow velocity,
$\overline{v^z/v^\mathrm{sat}}$ (green).
This figure shows that
the maximum velocity is about $0.1-0.4$,
and it is not so sensitive to the mass accretion rate.
We also find that 
$\overline{v^z/v^\mathrm{sat}}$
is very close to unity
independently of the mass accretion rate,
although we find 
$\overline{v^z/v^\mathrm{sat}} \lesssim 1$
at the rage of $\dot{M}_{\rm BH}\lesssim 10^{2.5}L_{\rm E}$.
It means that the outflow velocity is determined by 
the force balance between the radiation drag and the radiation flux force.
The maximum velocity of the radiatively driven outflow/jet
from the super-critical accretion disks
cannot largely exceed a few 10\% of the light velocity.

The saturation velocity given in equation (\ref{eq:vsat}) is obtained by
assuming that 
$|v^z P_\mathrm{r}^{zz}| \gg  |v^R P_\mathrm{r}^{Rz}|, 
|v^\phi P_\mathrm{rad}^{\phi z}|$. This assumption comes 
from the fact that the diagonal terms of the radiation pressure dominates 
over the off-diagonal terms. This result does not change even if we apply the 
Eddington approximation instead of M-1 closure. We calculated the off-diagonal 
terms using the Eddington approximation, and confirmed that the diagonal 
terms are larger than off-diagonal terms. The difference of the radiation
dragging force is less than 1\%, implying that the saturation velocity does not 
change so much. It is, however, noted that the radiation pressure tensor in
the Eddington approximation is computed based on 
$E_\mathrm{rad}, F_\mathrm{rad}^i$, and $u^i$ obtained 
from the present simulation. If the numerical simulations with the Eddington 
approximation is performed, we may obtain the different results since the
Eddington approximation gives the different radiation flux from the
M-1 method (\citealt{2013ApJ...772..127T}; see also Section \ref{sec:summary}).

Here it seems that the terminal velocity approaches to $0.5$ in the free 
streaming limit 
(e.g., $F_\mathrm{rad}^z = E_\mathrm{rad}, P_\mathrm{rad}^{zz} =
E_\mathrm{rad}$) according to equation (\ref{eq:vsat}). 
This is because we ignored the relativistic correction, $f^j_\mathrm{rad-corr}$. 
If we take into account this term correctly, and if we assume 
$F_\mathrm{rad}^z=E_\mathrm{rad}$ (and
$P_\mathrm{rad}^{zz}=E_\mathrm{rad}$), 
the terminal velocity approaches to the light speed. 
Although we have $F_\mathrm{rad}^z < E_\mathrm{rad}$ and $v\ll 1$ in the
present simulations since the 
radiation comes from the funnel-shaped photosphere of the disk,
$F_\mathrm{rad}^z$ does 
approach to $E_\mathrm{rad}$ at very large altitude. But then, the
radiation is attenuated and the gas would not be effectively accelerated.

\begin{figure}
 \includegraphics[width=8cm]{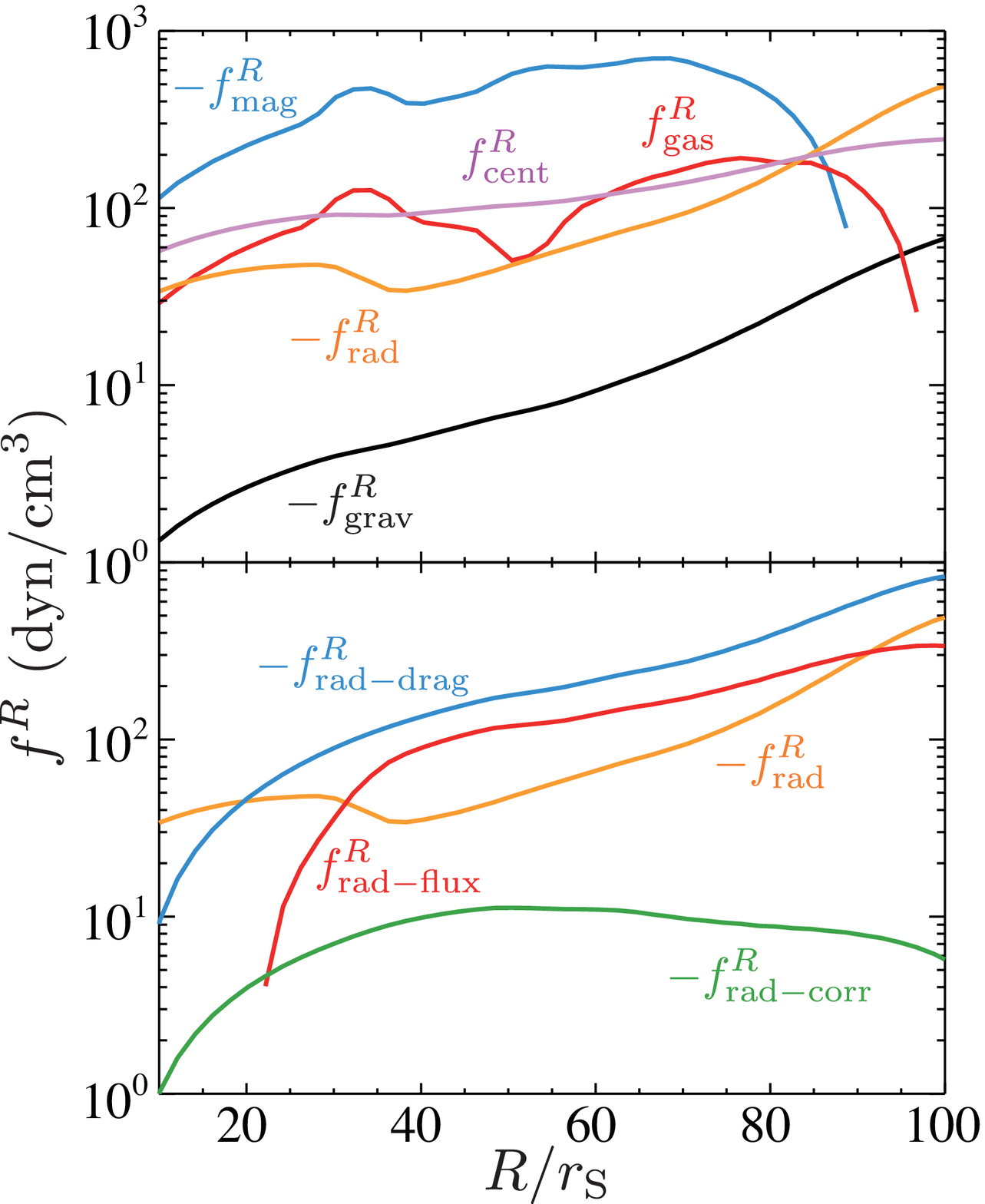}
  \caption{
 Horizontal component of the force densities 
 at $z=300 r_\mathrm{S}$ as a function of $R$.
 Top panel: the gravity ($-f^R_\mathrm{grav}$, black), 
 the radiation force ($-f^R_\mathrm{rad}$, orange), 
 the gas pressure gradient force ($f^R_\mathrm{gas}$, red), 
 the centrifugal force ($f^R_\mathrm{cent}$, magenta), 
 and the Lorentz force ($-f^R_\mathrm{mag}$, blue).
 Bottom panel: 
 The radiation flux force ($f^R_\mathrm{rad-flux}$, red),
 the radiation drag force ($-f^R_\mathrm{rad-drag}$, blue), 
 and the relativistic correction ($-f^R_\mathrm{rad-corr}$, green),
 The orange curve is the same as that of the top panel
 ($-f^R_\mathrm{rad}$).}
\label{fig:fig8}
\end{figure}


In cooperation with the Lorentz force,
the radiation drag also plays an important role for the 
collimation of the outflow.
A top panel of figure \ref{fig:fig8} shows the horizontal 
component of the force densities at $z=300 r_{\rm S}$.
Black, orange, red, magenta, and blue curves denote for the force
densities due to the gravitational force, the radiation force, 
the gas pressure gradient force, the centrifugal force, 
and the Lorentz force.
At the regime of $R\lesssim 100r_{\rm S}$,
the gravitational, the radiation and 
the Lorentz forces are negative (i.e., $-R$ direction), while 
the gas pressure gradient and centrifugal forces are positive.
The sum of $f^R_\mathrm{gas}$ and $f^R_\mathrm{cent}$
approximately balances with the Lorentz force. 
This implies that 
the gas pressure gradient and centrifugal forces 
work to expand the outflow in $R$-direction,
in contrast that the Lorentz force collimates the outflow.
Such a result is similar with 
the result of non-relativistic RMHD simulations 
\citep{2010PASJ...62L..43T},
in which the centrifugal force balances with the Lorentz force.
At the far region, $R\gtrsim 80r_{\rm S}$,
the radiation force has a negative value
and is dominant over the other forces.

Negative value of the radiation force is responsible for
the radiation drag.
It is clearly seen in a bottom panel of Figure
\ref{fig:fig8}. 
In this panel, 
we plot $-f_\mathrm{rad}^R$ (orange), 
$f^R_\mathrm{rad-flux}$ (red), 
$-f^R_\mathrm{rad-drag}$ (blue), 
and $-f^R_\mathrm{rad-corr}$ (green) at $z=300 r_{\rm S}$.
The radiation drag force dominates over the radiation flux force 
and the relativistic correction except at $R\leq 20r_{\rm S}$,
where the inward ($-R$ direction) radiation flux force is dominant.
Hence, 
the expansion of the outflow is prevented by the radiation drag
at $R\gtrsim 80r_{\rm S}$.
We conclude that 
the radiation drag force has important roles for determining both
the velocity and the opening angle of the outflow.



\section{Conclusions and Discussion}\label{sec:summary}
We performed 2.5-dimensional special relativistic radiation 
magnetohydrodynamics simulations to study
the super-critical accretion disks and 
the radiatively driven outflows.
We found that the outflow 
is accelerated by the radiation flux force 
and is subjected to the radiation drag force,
which prevents the outflow from speeding up.
The outflow velocity 
is determined by force balance between above two forces
and becomes $30-40 \%$ of the light speed
near the rotation axis.
Such a velocity does not change so much 
in the super-critical accretion regime
of $\dot{M}_{\rm BH} \sim 10^2-10^3 L_{\rm Edd}$.

Such a faster outflow is surrounded by a slower outflow, 
of which the velocity is $\sim 0.1$.
It is similar with so-called spine-sheath structure.
It is found that the radiation drag force 
works to collimate the slower outflow
in cooperation with the Lorentz force,
though the faster outflow is mainly collimated by the Lorentz force.
The most of kinetic energy is carried by the slower outflow,
and the radiation energy is larger in 
the slower outflow than in the faster outflow.
Although the mass accretion rate onto the black hole
largely exceeds the critical rate ($L_{\rm Edd}$),
the total photon and kinematic luminosities are comparable to $L_E$,
since a huge amount of the radiation energy is trapped 
and swallowed by the black hole with accreting matter.
The swallowed photon luminosity is roughly proportional to 
the mass accretion rate.

The resulting outflow velocity of $\sim 0.3-0.4$
nicely agrees with the jets ($\sim 0.26$) observed in SS 433
\citep{2002ApJ...564..941M}.
However, it is inconsistent with the highly 
relativistic jets in GRS 1915+195
\citep{1994Natur.371...46M,1999MNRAS.304..865F}, or some active galactic
nuclei \citep{1995ApJ...447..582B,2012A&A...538L..10G}.
The outflow velocity might increase
if the slower outflow is optically thick.
This is because the outflow is self-shielded 
and the spine of the outflow can avoid suffering from the radiation drag.
In addition, if the photon bubble instability occurs
near the innermost regions of the accretion disks,
the outward radiation flux increases,
going up the outflow velocity due to the enhanced radiation flux force
\citep{1978MNRAS.184...53B, 2005ApJ...624..267T}.
On the other hand, the outflow velocity does not increase 
even if the pair-plasma appears in the outflow, 
since the radiation drag acts on the positron as well as the electron.

We included only opacities due to the electron scattering and 
free-free absorption. Recently, \cite{2009PASJ...61..769K} performed 
non-relativistic radiation hydrodynamic simulations including thermal 
Comptonization. They showed that although the Compton effect does not impact 
on the disk structure so much, the gas temperature of the outflow drastically 
decreases due to the Compton up-scattering. Decreasing of the gas temperature 
leads to softening of X-ray spectrum \citep{2012ApJ...752...18K}. We need to 
include the Comptonization in relativistic radiation MHD simulation 
\cite{2014arXiv1407.4421S} to explain the spectral properties observed in 
microquasars.

Although we assume the axisymmetry in the present study,
three-dimensional study is required in order to
estimate the outflow velocity more precisely.
If the density of the disk fluctuates in azimuthal direction,
the photon trapping effect might degrade since 
the photon can escape from the less dense region.
Then,
the radiation fields change and 
the outflow velocity alters.
The dynamo would change the evolution of the magnetic fields,
leading to the change of the disk structure. 
The saturation velocity in such a case 
might be differ from that of the present study.
The 3-dimensional simulation has been initiated
in the Newtonian limit \citep{2014arXiv1410.0678J} and 
in general relativity \citep{2014MNRAS.441.3177M}.
\cite{2014arXiv1410.0678J} performed 3-dimensional simulation of slim
disks by solving full radiation transfer equation keeping terms of
$\mathcal{O}(v)$. They found that the
outflow velocity is about 0.3, which is consistent with our results. 
Comparing results in detail between these different models is also very important
future work to
understand the validity of the closure relation.

The light bending should decrease the photon luminosity,
since more photons are swallowed by the black hole.
Then, the kinetic luminosity of the outflow would go down.
In contrast, 
the spin of the black hole is though to enhance 
the outflow (e.g., \citeauthor{1977MNRAS.179..433B}
\citeyear{1977MNRAS.179..433B}).
Recently, GR simulations have been revealed 
that the strong outflow is generated 
from the super-critical accretion disks 
around Kerr black hole \citep{2014MNRAS.439..503S, 2014MNRAS.441.3177M}.

Finally, we discuss the difference between our simulations and previous 
non-relativistic simulations. \cite{2010PASJ...62L..43T} performed non-relativistic
radiation MHD simulation with the flux-limited diffusion approximation (FLD, 
\citet{1981ApJ...248..321L}), reporting that the outflow velocity is about 
$0.6-0.7$, which is faster than our results ($\sim 0.4$). Such a
difference would 
be caused by the radiation drag effect, which is not taken into account
in \cite{2010PASJ...62L..43T}. 
Thus, we stress that it is important to include the radiation
drag effects to study acceleration mechanisms of outflows from black hole
accretion disks.

The different algorithms make differences especially in the outflow regions,
since the approximate radiative transfer algorithms (e.g., FLD approximation, 
the Eddington approximation, and the M-1 method) are known to be problematic 
in the optically thin region and, in contrast, give accurate radiation fields 
in the optically thick diffusion limit. The FLD approximation cannot be 
properly applied to the relativistic fluid since it violates a causality 
(e.g., the radiation energy equation becomes parabolic). On the other hand, 
the Eddington approximation is utilizable in relativistic simulations. In the 
relativistic one-dimensional test problems, \citep{2013ApJ...772..127T} showed 
that $F_\mathrm{rad}'/E_\mathrm{rad}'$ becomes larger for the M-1
treatment than for the Eddington approximation. Thus, it is expected that the 
terminal velocity becomes slower if the Eddington approximation is adopted. 
However, the performing simulation with the Eddington approximation is time-consuming 
since $6\times 6$ matrix inversion at each grid point is needed to
compute Eddington tensor. Hence, 
the detailed comparison between Eddington approximation and M-1 closure is 
beyond the scope of the present paper and is left as an important future 
work. In addition, recently, a more accurate method, which solves radiative 
transfer equation, has been proposed by \cite{2014ApJS..213....7J}. 
Comparing results between these different models is also
very important future work to understand the validity of the closure relation.
\bigskip
We are grateful to an anonymous referee for improving our manuscript.
We thank T. Kawashima for useful discussion.
Numerical computations were carried out on Cray XC30 at the Center for
Computational Astrophysics, CfCA, at the National Astronomical Observatory
of Japan, and on T2K at the University of
Tokyo. This work is supported in part by Ministry of Education, Culture,
Sports, Science, and Technology (MEXT) for Young Scientist (B) 24740127 (K.O.). 
A part of this research has been funded by MEXT HPCI STRATEGIC PROGRAM
and the Center for the Promotion of Integrated Sciences (CPIS) of Sokendai.


\begin{thebibliography}{45}
\expandafter\ifx\csname natexlab\endcsname\relax\def\natexlab#1{#1}\fi

\bibitem[{{Abdo} {et~al.}(2009){Abdo}, {Ackermann}, {Arimoto}, {Asano},
  {Atwood}, {Axelsson}, {Baldini}, {Ballet}, {Band}, {Barbiellini}, \&
  et~al.}]{2009Sci...323.1688A}
{Abdo}, A.~A., {Ackermann}, M., {Arimoto}, M., {Asano}, K., {Atwood}, W.~B.,
  {Axelsson}, M., {Baldini}, L., {Ballet}, J., {Band}, D.~L., {Barbiellini},
  G., \& et~al. 2009, Science, 323, 1688

\bibitem[{{Abramowicz} {et~al.}(1988){Abramowicz}, {Czerny}, {Lasota}, \&
  {Szuszkiewicz}}]{1988ApJ...332..646A}
{Abramowicz}, M.~A., {Czerny}, B., {Lasota}, J.~P., \& {Szuszkiewicz}, E. 1988,
  \apj, 332, 646

\bibitem[{{Begelman}(1978)}]{1978MNRAS.184...53B}
{Begelman}, M.~C. 1978, \mnras, 184, 53

\bibitem[{{Biretta} {et~al.}(1995){Biretta}, {Zhou}, \&
  {Owen}}]{1995ApJ...447..582B}
{Biretta}, J.~A., {Zhou}, F., \& {Owen}, F.~N. 1995, \apj, 447, 582

\bibitem[{{Blandford} \& {Znajek}(1977)}]{1977MNRAS.179..433B}
{Blandford}, R.~D. \& {Znajek}, R.~L. 1977, \mnras, 179, 433

\bibitem[{{Bucciantini} \& {Del Zanna}(2013)}]{2013MNRAS.428...71B}
{Bucciantini}, N. \& {Del Zanna}, L. 2013, \mnras, 428, 71

\bibitem[{{Bugli} {et~al.}(2014){Bugli}, {Del Zanna}, \&
  {Bucciantini}}]{2014MNRAS.440L..41B}
{Bugli}, M., {Del Zanna}, L., \& {Bucciantini}, N. 2014, \mnras, 440, L41

\bibitem[{{Cowling}(1933)}]{1933MNRAS..94...39C}
{Cowling}, T.~G. 1933, \mnras, 94, 39

\bibitem[{{Farris} {et~al.}(2008){Farris}, {Li}, {Liu}, \&
  {Shapiro}}]{2008PhRvD..78b4023F}
{Farris}, B.~D., {Li}, T.~K., {Liu}, Y.~T., \& {Shapiro}, S.~L. 2008, \prd, 78,
  024023

\bibitem[{{Fender} {et~al.}(1999){Fender}, {Garrington}, {McKay}, {Muxlow},
  {Pooley}, {Spencer}, {Stirling}, \& {Waltman}}]{1999MNRAS.304..865F}
{Fender}, R.~P., {Garrington}, S.~T., {McKay}, D.~J., {Muxlow}, T.~W.~B.,
  {Pooley}, G.~G., {Spencer}, R.~E., {Stirling}, A.~M., \& {Waltman}, E.~B.
  1999, \mnras, 304, 865

\bibitem[{{Giroletti} {et~al.}(2012){Giroletti}, {Hada}, {Giovannini},
  {Casadio}, {Beilicke}, {Cesarini}, {Cheung}, {Doi}, {Krawczynski}, {Kino},
  {Lee}, \& {Nagai}}]{2012A&A...538L..10G}
{Giroletti}, M., {Hada}, K., {Giovannini}, G., {Casadio}, C., {Beilicke}, M.,
  {Cesarini}, A., {Cheung}, C.~C., {Doi}, A., {Krawczynski}, H., {Kino}, M.,
  {Lee}, N.~P., \& {Nagai}, H. 2012, \aap, 538, L10

\bibitem[{{Hawley} \& {Balbus}(1991)}]{1991ApJ...376..223H}
{Hawley}, J.~F. \& {Balbus}, S.~A. 1991, \apj, 376, 223

\bibitem[{{Ichimaru}(1977)}]{1977ApJ...214..840I}
{Ichimaru}, S. 1977, \apj, 214, 840

\bibitem[{{Igumenshchev}(2008)}]{2008ApJ...677..317I}
{Igumenshchev}, I.~V. 2008, \apj, 677, 317

\bibitem[{{Igumenshchev} {et~al.}(2003){Igumenshchev}, {Narayan}, \&
  {Abramowicz}}]{2003ApJ...592.1042I}
{Igumenshchev}, I.~V., {Narayan}, R., \& {Abramowicz}, M.~A. 2003, \apj, 592,
  1042

\bibitem[{{Jiang} {et~al.}(2014{\natexlab{a}}){Jiang}, {Stone}, \&
  {Davis}}]{2014arXiv1410.0678J}
{Jiang}, Y.-F., {Stone}, J.~M., \& {Davis}, S.~W. 2014{\natexlab{a}}, ArXiv
  e-prints

\bibitem[{{Jiang} {et~al.}(2014{\natexlab{b}}){Jiang}, {Stone}, \&
  {Davis}}]{2014ApJS..213....7J}
---. 2014{\natexlab{b}}, \apjs, 213, 7

\bibitem[{{Kato} {et~al.}(2008){Kato}, {Fukue}, \&
  {Mineshige}}]{2008bhad.book.....K}
{Kato}, S., {Fukue}, J., \& {Mineshige}, S. 2008, {Black-Hole Accretion Disks
  --- Towards a New Paradigm ---}, ed. {Kato, S., Fukue, J., \& Mineshige, S.}

\bibitem[{{Kawashima} {et~al.}(2009){Kawashima}, {Ohsuga}, {Mineshige},
  {Heinzeller}, {Takabe}, \& {Matsumoto}}]{2009PASJ...61..769K}
{Kawashima}, T., {Ohsuga}, K., {Mineshige}, S., {Heinzeller}, D., {Takabe}, H.,
  \& {Matsumoto}, R. 2009, \pasj, 61, 769

\bibitem[{{Kawashima} {et~al.}(2012){Kawashima}, {Ohsuga}, {Mineshige},
  {Yoshida}, {Heinzeller}, \& {Matsumoto}}]{2012ApJ...752...18K}
{Kawashima}, T., {Ohsuga}, K., {Mineshige}, S., {Yoshida}, T., {Heinzeller},
  D., \& {Matsumoto}, R. 2012, \apj, 752, 18

\bibitem[{{Levermore} \& {Pomraning}(1981)}]{1981ApJ...248..321L}
{Levermore}, C.~D. \& {Pomraning}, G.~C. 1981, \apj, 248, 321

\bibitem[{{Marshall} {et~al.}(2002){Marshall}, {Canizares}, \&
  {Schulz}}]{2002ApJ...564..941M}
{Marshall}, H.~L., {Canizares}, C.~R., \& {Schulz}, N.~S. 2002, \apj, 564, 941

\bibitem[{{McKinney} {et~al.}(2014){McKinney}, {Tchekhovskoy}, {Sadowski}, \&
  {Narayan}}]{2014MNRAS.441.3177M}
{McKinney}, J.~C., {Tchekhovskoy}, A., {Sadowski}, A., \& {Narayan}, R. 2014,
  \mnras, 441, 3177

\bibitem[{{Meier}(2003)}]{2003NewAR..47..667M}
{Meier}, D.~L. 2003, \nar, 47, 667

\bibitem[{{Mirabel} \& {Rodr{\'{\i}}guez}(1994)}]{1994Natur.371...46M}
{Mirabel}, I.~F. \& {Rodr{\'{\i}}guez}, L.~F. 1994, \nat, 371, 46

\bibitem[{{Narayan} \& {Yi}(1994)}]{1994ApJ...428L..13N}
{Narayan}, R. \& {Yi}, I. 1994, \apjl, 428, L13

\bibitem[{{Ohsuga} \& {Mineshige}(2007)}]{2007ApJ...670.1283O}
{Ohsuga}, K. \& {Mineshige}, S. 2007, \apj, 670, 1283

\bibitem[{{Ohsuga} \& {Mineshige}(2011)}]{2011ApJ...736....2O}
---. 2011, \apj, 736, 2

\bibitem[{{Ohsuga} {et~al.}(2009){Ohsuga}, {Mineshige}, {Mori}, \&
  {Kato}}]{2009PASJ...61L...7O}
{Ohsuga}, K., {Mineshige}, S., {Mori}, M., \& {Kato}, Y. 2009, \pasj, 61, L7+

\bibitem[{{Ohsuga} {et~al.}(2002){Ohsuga}, {Mineshige}, {Mori}, \&
  {Umemura}}]{2002ApJ...574..315O}
{Ohsuga}, K., {Mineshige}, S., {Mori}, M., \& {Umemura}, M. 2002, \apj, 574,
  315

\bibitem[{{Ohsuga} {et~al.}(2003){Ohsuga}, {Mineshige}, \&
  {Watarai}}]{2003ApJ...596..429O}
{Ohsuga}, K., {Mineshige}, S., \& {Watarai}, K.-y. 2003, \apj, 596, 429

\bibitem[{{Ohsuga} {et~al.}(2005){Ohsuga}, {Mori}, {Nakamoto}, \&
  {Mineshige}}]{2005ApJ...628..368O}
{Ohsuga}, K., {Mori}, M., {Nakamoto}, T., \& {Mineshige}, S. 2005, \apj, 628,
  368

\bibitem[{{Paczy{\'n}sky} \& {Wiita}(1980)}]{1980A&A....88...23P}
{Paczy{\'n}sky}, B. \& {Wiita}, P.~J. 1980, \aap, 88, 23

\bibitem[{{Roedig} {et~al.}(2012){Roedig}, {Zanotti}, \&
  {Alic}}]{2012MNRAS.426.1613R}
{Roedig}, C., {Zanotti}, O., \& {Alic}, D. 2012, \mnras, 426, 1613

\bibitem[{{Rykoff} {et~al.}(2009){Rykoff}, {Aharonian}, {Akerlof}, {Ashley},
  {Barthelmy}, {Flewelling}, {Gehrels}, {G{\"o}{\v g}{\"u}{\c s}}, {G{\"u}ver},
  {Kizilo{\v g}lu}, {Krimm}, {McKay}, {{\"O}zel}, {Phillips}, {Quimby},
  {Rowell}, {Rujopakarn}, {Schaefer}, {Smith}, {Vestrand}, {Wheeler}, {Wren},
  {Yuan}, \& {Yost}}]{2009ApJ...702..489R}
{Rykoff}, E.~S., {Aharonian}, F., {Akerlof}, C.~W., {Ashley}, M.~C.~B.,
  {Barthelmy}, S.~D., {Flewelling}, H.~A., {Gehrels}, N., {G{\"o}{\v g}{\"u}{\c
  s}}, E., {G{\"u}ver}, T., {Kizilo{\v g}lu}, {\"U}., {Krimm}, H.~A., {McKay},
  T.~A., {{\"O}zel}, M., {Phillips}, A., {Quimby}, R.~M., {Rowell}, G.,
  {Rujopakarn}, W., {Schaefer}, B.~E., {Smith}, D.~A., {Vestrand}, W.~T.,
  {Wheeler}, J.~C., {Wren}, J., {Yuan}, F., \& {Yost}, S.~A. 2009, \apj, 702,
  489

\bibitem[{{Sadowski} {et~al.}(2014){Sadowski}, {Narayan}, {Tchekhovskoy},
  {Abarca}, {Zhu}, \& {McKinney}}]{2014arXiv1407.4421S}
{Sadowski}, A., {Narayan}, R., {Tchekhovskoy}, A., {Abarca}, D., {Zhu}, Y., \&
  {McKinney}, J.~C. 2014, ArXiv e-prints

\bibitem[{{S{\c a}dowski} {et~al.}(2014){S{\c a}dowski}, {Narayan}, {McKinney},
  \& {Tchekhovskoy}}]{2014MNRAS.439..503S}
{S{\c a}dowski}, A., {Narayan}, R., {McKinney}, J.~C., \& {Tchekhovskoy}, A.
  2014, \mnras, 439, 503

\bibitem[{{S{\c a}dowski} {et~al.}(2013){S{\c a}dowski}, {Narayan},
  {Tchekhovskoy}, \& {Zhu}}]{2013MNRAS.429.3533S}
{S{\c a}dowski}, A., {Narayan}, R., {Tchekhovskoy}, A., \& {Zhu}, Y. 2013,
  \mnras, 429, 3533

\bibitem[{{Shakura} \& {Sunyaev}(1973)}]{1973A&A....24..337S}
{Shakura}, N.~I. \& {Sunyaev}, R.~A. 1973, \aap, 24, 337

\bibitem[{{Sol} {et~al.}(1989){Sol}, {Pelletier}, \&
  {Asseo}}]{1989MNRAS.237..411S}
{Sol}, H., {Pelletier}, G., \& {Asseo}, E. 1989, \mnras, 237, 411

\bibitem[{{Takahashi} \& {Ohsuga}(2013)}]{2013ApJ...772..127T}
{Takahashi}, H.~R. \& {Ohsuga}, K. 2013, \apj, 772, 127

\bibitem[{{Takahashi} {et~al.}(2013){Takahashi}, {Ohsuga}, {Sekiguchi},
  {Inoue}, \& {Tomida}}]{2013ApJ...764..122T}
{Takahashi}, H.~R., {Ohsuga}, K., {Sekiguchi}, Y., {Inoue}, T., \& {Tomida}, K.
  2013, \apj, 764, 122

\bibitem[{{Takeuchi} {et~al.}(2010){Takeuchi}, {Ohsuga}, \&
  {Mineshige}}]{2010PASJ...62L..43T}
{Takeuchi}, S., {Ohsuga}, K., \& {Mineshige}, S. 2010, \pasj, 62, L43+

\bibitem[{{Turner} {et~al.}(2005){Turner}, {Blaes}, {Socrates}, {Begelman}, \&
  {Davis}}]{2005ApJ...624..267T}
{Turner}, N.~J., {Blaes}, O.~M., {Socrates}, A., {Begelman}, M.~C., \& {Davis},
  S.~W. 2005, \apj, 624, 267

\bibitem[{{Zanotti} {et~al.}(2011){Zanotti}, {Roedig}, {Rezzolla}, \& {Del
  Zanna}}]{2011MNRAS.417.2899Z}
{Zanotti}, O., {Roedig}, C., {Rezzolla}, L., \& {Del Zanna}, L. 2011, \mnras,
  417, 2899

\end{thebibliography}

\end{document}